\documentstyle[12pt,preprint,aps]{revtex}

\begin{document}
\title
{Andreev reflection in engineered Al/Si/InGaAs(001) junctions 
}
\author{Silvano De Franceschi, Francesco Giazotto, 
and Fabio Beltram}
\address{Scuola Normale Superiore and INFM, I-56126 Pisa,
Italy}
\author{Lucia Sorba, Marco Lazzarino, and Alfonso Franciosi$^{a)}$}
\address{Laboratorio Nazionale TASC-INFM, Area di Ricerca, Padriciano
99, I-34012 Trieste, Italy}
\footnotetext[1]{Also with Universit\`a di Trieste, Trieste, Italy.}
\author{To be published in Phyl. Mag. B}
\maketitle

\begin{abstract}

Complete suppression of the native n-type Schottky barrier is demonstrated in Al/InGaAs(001) junctions grown by molecular-beam-epitaxy. This result was achieved by the insertion of Si bilayers at the metal-semiconductor interface allowing the realization of truly Ohmic non-alloyed contacts in low-doped and low-In content InGaAs/Si/Al junctions. It is shown that this technique is ideally suited for the fabrication of high-transparency superconductor-semiconductor junctions. To this end magnetotransport characterization of Al/Si/InGaAs low-n-doped single junctions below the Al critical temperature is presented. Our measurements show Andreev-reflection dominated transport corresponding to junction transparency close to the theoretical limit due to Fermi-velocity mismatch.

\end{abstract}


In the last few years there has been an increasing interest in the study of
semiconductor-superconductor (Sm-S) hybrid systems \cite{been,klein,lamb}. These allow
the investigation of exotic coherent-transport effects and have great 
potential for device applications.  
The characteristic physical phenomenon driving electron transport at a 
S-Sm junction is Andreev reflection \cite{andreev}. 
In this process (originally observed in normal metal-superconductor junctions
) an electron incident from the Sm side  on the superconductor 
may be transmitted as a Cooper pair if a hole is retroreflected 
along the time-reversed path of the incoming particle.  
High junction transparency is a crucial property for the observation of 
Andreev-reflection dominated transport. 
Different techniques have been explored to meet this requirement   
including metal deposition immediately after 
As-decapping \cite{kast}, 
Ar$^+$ back-sputtering \cite{nguy}, and {\it in situ} metallization in 
the molecular-beam epitaxy (MBE) chamber \cite{akaz}. All these tests were performed in InAs-based Sm-S devices where the main transmittance-limiting factor is interface contamination. On the contrary, for semiconductor materials such as those grown on either GaAs or InP, the strongest limitation arises from the presence of a native Schottky barrier. In this case, in order to enhance junction transparency  penetrating contacts \cite{gao,williams} and heavily doped surface layers 
\cite{kast,tabo} were used.
Recently  we have reported on a new technique \cite{silv}, 
alternative to doping, to obtain Schottky-barrier-free 
Al/n-In$_x$Ga$_{1-x}$As(001) junctions ($x \agt 0.3$) by MBE growth.
This is based on the inclusion of an ultrathin Si interface layer 
under As flux which 
changes the pinning position of the Fermi level at the metal-semiconductor 
junction and leads to the total suppression of the Schottky barrier.
In this work we show the behavior of such Ohmic contacts realized and furthermore demonstrate how this method can be successfully exploited to obtain high transparency Sm-S hybrid junctions \cite{franz}. Notably these are based on low-doped and low-In-content InGaAs alloys that are ideal 
candidates for the implementation of ballistic-transport structures. 

Al/n-In$_{0.38}$Ga$_{0.62}$As junctions incorporating Si 
interface layers were 
grown by MBE. Their schematic structure 
is shown in Fig. 1.  The semiconductor portion consists of a 
300-nm-thick GaAs buffer layer grown at 600 $^\circ$C on n-type GaAs(001) 
and Si-doped  at $n \sim 10^{18}$ cm$^{-3}$ followed by 
a 2-$\mu$m-thick n-In$_{0.38}$Ga$_{0.62}$As layer grown at
500 $^\circ$C with an inhomogeneous doping profile. 
The top 1.5-$\mu$m-thick region was doped at $n=6.5 \cdot 10^{16}$ 
cm$^{-3}$, 
the bottom buffer region (0.5 $\mu$m thick) was
heavily doped at $n \sim 10^{18}$ cm$^{-3}$.
After In$_{0.38}$Ga$_{0.62}$As growth the substrate temperature
was lowered to 300$^\circ$C and a Si atomic bilayer was deposited
under As flux \cite{silv}.
Al deposition ($\simeq 150$ nm) was carried out {\it in situ} at room temperature. During Al deposition  the pressure in the MBE
chamber was below $5 \cdot 10^{-10}$ Torr.
Reference Al/n-In$_{0.38}$Ga$_{0.62}$As junctions were also grown 
with the same semiconductor part but without the Si interface layer.

In order to compare the current-voltage ($I$--$V$) behavior of 
Si-engineered and reference junctions,  
circular contacts were defined on the top surface 
with various diameters in the 
75--150 $\mu$m range. 
 Standard photolithographic techniques and wet chemical etching were used to
this aim.
Back contacting was provided for electrical characterization by
metallizing the whole substrate bottom. 
$I$--$V$ characterization was performed 
in the 20--300 K
temperature range using a closed-cycle cryostat equipped with microprobes.  
Typical room-temperature (dashed lines) and low-temperature (solid lines) 
$I$--$V$ characteristics for both Si-engineered and reference diodes 
are shown in Fig. 2.

The reference diode exhibits a marked rectifying behavior which is 
enhanced at low temperatures. 
We have measured the corresponding barrier height  by different techniques:
thermionic-emission $I$--$V$ measurements in the 270--300 K temperature range,
and linear fit in the forward bias region of log($I$)--$V$ characteristics 
measured at $\sim 200$ K. These two approaches 
yielded barrier heights of $0.22 \pm 0.05$ eV and $0.23 \pm 0.02$ eV
respectively. These values include corrections    
for image-charge and thermionic-field-emission effects \cite{S}. 
The quoted uncertainties reflect diode to diode fluctuations and uncertainties 
in the barrier height  determination.      

The engineered diode shows no rectifying behavior even at low temperatures (20 K in Fig. 2). Its $I$--$V$ characteristics bear no trace of a SB and are linear over the whole 20--300 K
temperature range. Their slope is only weakly affected by temperature.  
To investigate the possible existence 
of a residual SB whose rectifying effect might be hidden by the
series resistance arising from the InGaAs 
bulk and the back contact, 
we modeled the low-temperature $I$--$V$ behavior of the engineered diode 
in terms of a residual barrier height  $\phi_n$ and a series resistance $R$ \cite{PS}.
We were able to reproduce the experimental $I$--$V$ curves only with   
$\phi_n < 0.03$ eV. As will be apparent from what follows, this value represents only an upper limit for the barrier height.

Doping effects do not play any significant role in
the barrier suppression. In order to verify this, we annealed the engineered diode  at 420
$^\circ$C for 5 seconds. 
Following this we observed a marked rectifying behavior analogous to that of the reference sample. This result is in
line with the findings reported in Ref. \cite{S6B} on the  
thermal stability of Si-engineered SBs in Al/GaAs junctions and reflects Si redistribution at the interface. 
Wear-out tests on engineered diodes were also performed in order to
verify the persistence of the ohmic behavior against prolonged 
high-current stress. 
To this end we monitored the $I$--$V$ characteristics during  
24 hours of continuous operation at current densities 
of 200 A/cm$^2$. 
No changes were detected.

In order to demonstrate the applicability of this technique to the realization of  high transparency Sm-S hybrid devices, rectangular 100$\times$160 $\mu$m$^2$ Al/n-In$_{0.38}$Ga$_{0.62}$As 
junctions were patterned on the sample surface using standard 
photolithographic techniques and wet chemical etching. 
Two additional 100$\times$50 $\mu$m$^2$-wide and 200-nm-thick Au pads
were electron-beam evaporated on top of every Al pattern in order to allow
four-wire electrical measurements.  
Samples were mounted on non-magnetic dual-in-line sample
holders, and 25-$\mu$m-thick gold
wires were bonded to the gold pads.
$I$--$V$ characterizations as a function of temperature
($T$) and static magnetic field ($H$)   
were performed in a $^3$He closed-cycle cryostat. 

The critical temperature ($T_c$) of the Al film was 1.1 K (which corresponds to a gap $\Delta \approx 0.16$ meV).  
The normal-state resistance $R_N$ of our devices was 0.2 $\Omega$,
including the series-resistance contribution ($\approx 0.1 \Omega$) 
of the semiconductor. At $H=0$ and  below $T_c$, 
dc $I$--$V$ characteristics exhibited important non-linearities
around zero bias that can be visualized by plotting  the 
differential conductance ($G$) as a function of the applied bias ($V$).
In Fig. 3(a) we show a tipical set of $G$--$V$ curves obtained at
different temperatures in the 0.33--1.03 K range. 
Notably even at $T=0.33$ K,
i.e. well below $T_c$, a high value of $G$ is observed at zero bias. 
At low temperature and bias (i.e., when the voltage drop
across the junction is lower than $\Delta /e$ \cite{nota}), 
transport is dominated by Andreev reflection. 
The observation of such pronounced Andreev reflection demonstrates
high junction transparency. The latter can be quantified in terms of a  
dimensionless parameter $Z$ 
according to the Blonder-Tinkham-Klapwijk (BTK) model \cite{btk,z}. 
To analyze the data of Fig. 3(a) we followed the model 
by Chaudhuri and Bagwell \cite{chau},  which is the three-dimensional
generalization of the BTK model.
For our S-Sm junction we found $Z \approx 1$ corresponding
to a $\sim$50 \% normal-state transmission coefficient.  
We note that without the aid of the 
Si-interface-layer technique, doping concentrations 
over two orders of magnitude greater than that employed here 
would be necessary to achieve comparable transmissivity 
(see e.g. Refs. \cite{kast,gao,tabo,tabo2}).
This drastic reduction in the impurity concentration is a very
attractive feature for the fabrication of ballistic structures.     
It should also be noted that our reported $Z$
value is close to the intrinsic transmissivity limit 
related to the Fermi-velocity mismatch between Al and InGaAs
\cite{BT}. 

We should also like to comment on the homogeneity of our junctions. By applying the BTK formalism, 
$Z\approx 1$ leads to a  
calculated value of the normal-state resistance ($R_N^{th}$)
much smaller than the experimental value $R_N^{exp}$:
$R_N^{th}/R_N^{exp}=0.003$ . This would indicate that 
only a small fraction  ($R_N^{th}/R_N^{exp}$) of the contact area has the 
high transparency and leads to the transport properties of the junction,
as already reported for different fabrication
techniques \cite{gao,van}. Values of $R_N^{th}/R_N^{exp}$ ranging from
$\sim 10^{-4}$ to $\sim 10^{-2}$ can be found in the literature 
(see, e.g., Refs. \cite{kast,gao,van}).
Such estimates, however, should be taken with much caution. Experimentally, no homogeneities were observed on the lateral 
length scale of our contacts and we did observe a high uniformity in the transport 
properties of all junctions studied. 

The superconducting nature of the conductance dip for $|V|<\Delta/e$
is proved by its pronounced dependence on temperature and magnetic
field. Figure 3(a) shows how the zero-bias differential-conductance dip 
observed at $T=0.33$ K progressively weakens for $T$ approaching $T_c$.
This fact is consistent with the well-known
temperature-induced suppression of the superconducting energy gap 
$\Delta$. Far from $V=0$
the conductance is only marginally affected by temperature as
expected for a S-Sm junction when $|V|$ is significantly
larger than $\Delta/e$ \cite{btk}. 
A small depression in the zero-bias conductance is still observed at 
$T \simeq T_c$.  This, together with the slight asymmetry 
in the  $G$--$V$ curves, can be linked to a residual barrier  
at the buried InGaAs/GaAs heterojunction.

In Fig. 3(b) we show how the conductance can be strongly modified by 
very weak magnetic fields ($H$). The $G$--$V$ curves shown in Fig. 3(b) were
taken at $T=0.33$ K for different values of $H$ applied perpendicularly to the plane of the 
junction in the 
0--5 mT range. The superconducting gap vanishes for  
$H$ approaching the critical field ($H_c$) of 
the Al film ($H_c \simeq 10$ mT at $T=0.33$ K). 
Consequently, the zero-bias conductance dip is less and less
pronounced and at the same time shrinks with increasing magnetic field.
The latter effect was not as noticeable in Fig. 3(a) owing to the 
temperature-induced broadening of the single-particle Fermi distribution 
function \cite{btk}. 

In conclusion, we have reported on Ohmic behavior and Andreev-reflection dominated transport
in MBE-grown Si-engineered 
Al/n-In$_{0.38}$Ga$_{0.62}$As junctions. 
Transport properties
were studied as a function of temperature and magnetic field and 
showed junction transmissivity close to the theoretical limit for 
the S-Sm combination.
The present study demonstrates that the 
Si-interface-layer technique is a promising tool to 
obtain high-transparency S-Sm junctions involving InGaAs
alloys with low In content and low doping concentration.    
This technique yields Schottky-barrier-free junctions
without using InAs-based heterostructures and can be 
exploited in the most widespread MBE systems. It is 
particularly suitable for the realization of 
low-dimensional S-InGaAs hybrid systems grown on GaAs or InP substrates.
We should finally like to stress that its application in principle is not 
limited to Al metallizations and other superconductors 
could be equivalently used. In fact, to date the most convincing
interpretation of the silicon-assisted Schottky-barrier engineering is
based upon the heterovalency-induced IV/III-V local interface dipole
\cite{bin}. Within this description Schottky-barrier tuning is
a metal-independent effect.

The present work was supported by INFM under the PAIS project 
Eterostrutture Ibride Semiconduttore-Superconduttore and the TUSBAR program. One of us (F. G.) would like to acknowledge Europa Metalli 
S.p.A. for financial support. 


\begin{figure}
\caption{ 
Schematic structure of the Al/n-In$_{0.38}$Ga$_{0.62}$As junctions studied in
this work. Further details are given in the text.
} 
\end{figure}

\begin{figure}
\caption{Current-voltage characteristics of Si-engineered 
Al/Si/In$_{0.38}$Ga$_{0.62}$As diodes and their reference structure for 
100-$\mu$m-diameter devices.
Dashed and solid lines refer 
to room-temperature and low-temperature (20 K) measurements, respectively.
} 
\end{figure}

\begin{figure}
\caption{
(a) Differential conductance $vs$ bias of a Si-engineered Al/n-In$_{0.38}$Ga$_{0.62}$As single junction at several temperatures with no applied magnetic field.
(b) Differential conductance $vs$ bias of a Si-engineered Al/n-In$_{0.38}$Ga$_{0.62}$As single junction at $T=0.33$ K. Several characteristics under different  magnetic fields applied perpendicular to the 
junction plane are shown. 
}
\end{figure}


\begin{references}
\bibitem{note2} The values used for $m^*$ and $\epsilon$ were obtained by linear
interpolation between the corresponding InAs and GaAs parameters.
\bibitem{z} In this model $Z$ is related to the normal-state transmission coefficient  
$\Gamma$ by $\Gamma=(1+Z^2)^{-1}$.  
This approach was developed in the contex of ballistic systems, but
has been widely applied to diffusive systems (like the present one)
in order to gain an estimate of the junction trasmissivity 
\cite{gao,tabo,van}.
\end{references}
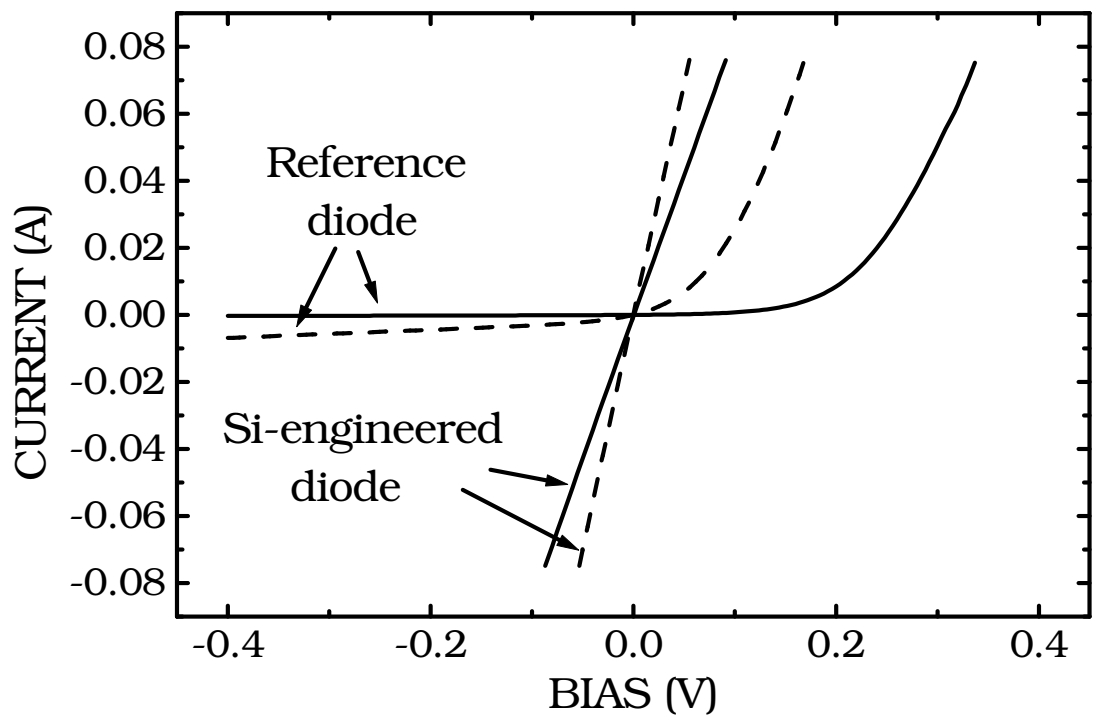
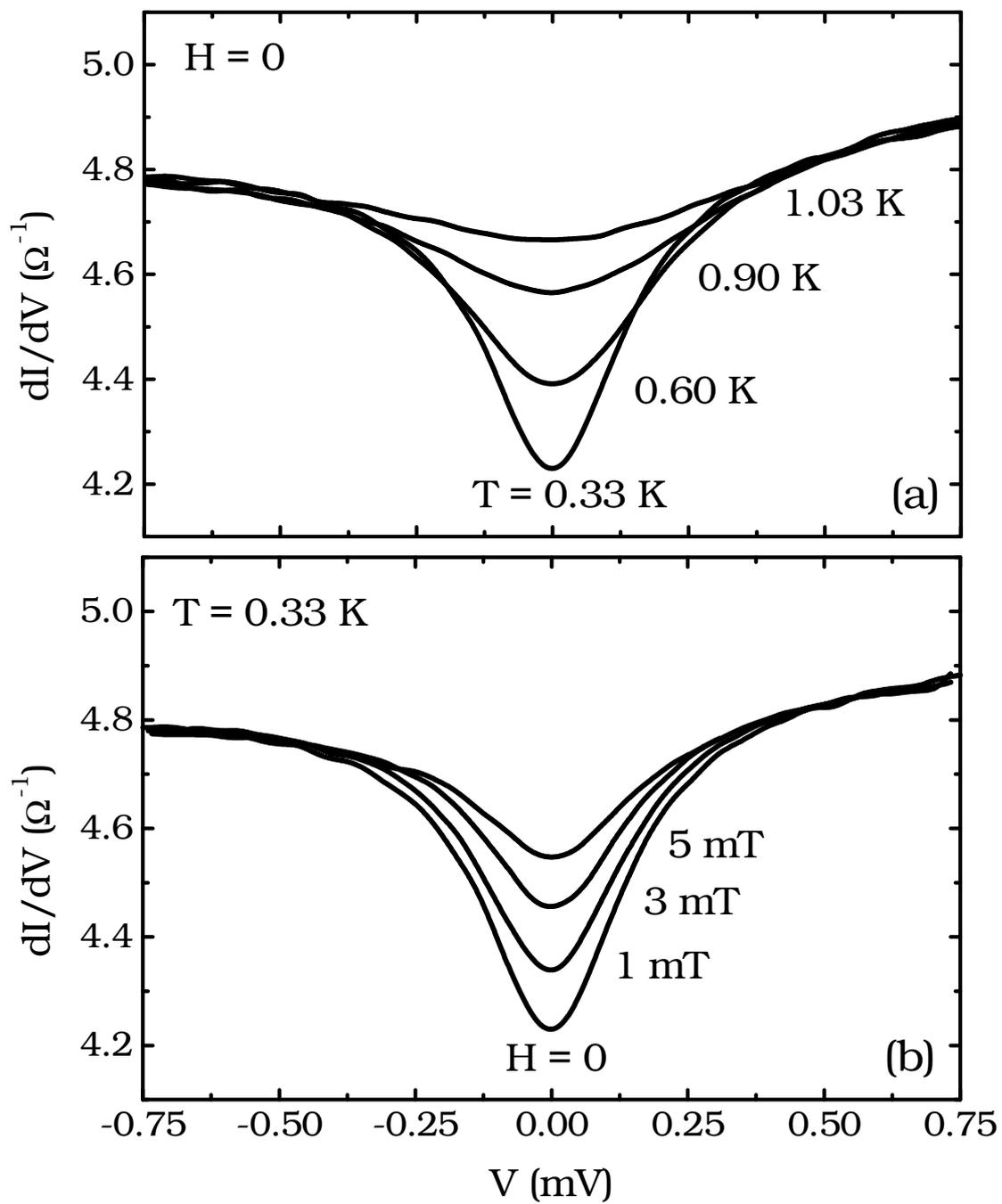
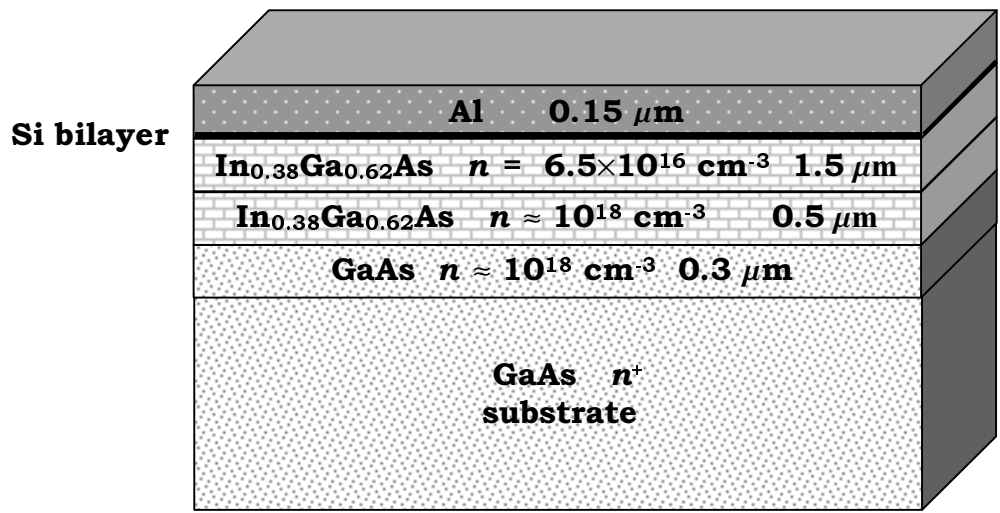
\end{document}